\documentclass[aps,prl,twocolumn,export]{revtex4-2} 
\usepackage[utf8]{inputenc}

\usepackage{graphicx}
\usepackage{fullpage}
\usepackage{tikz,tikz-cd} 
\usepackage{amsthm,amsmath,amssymb,amsfonts,mathtools,tensor,mhchem}

\usepackage{bbold} 

\usepackage[most]{tcolorbox}

\newcommand{\id}{\mathbb{1}}
\newcommand{\myinv}[1]{#1^{\scalebox{0.9}[1.0]{-}1}}
\newcommand{\m}{{\scalebox{0.9}[1.0]{-}1}}
\newcommand{\p}{{\scalebox{0.6}[0.6]{+1}}}

\newcommand{\tr}{{\rm Tr}}
\newcommand{\ket}[1]{| #1 \rangle}
\newcommand{\bra}[1]{\langle #1 |}

\renewcommand{\H}{\mathcal{H}}

\renewcommand{\P}{\mathcal{P}}
\newcommand{\G}{\mathcal{G}}

\usetikzlibrary{decorations.pathreplacing,decorations.markings,calc,3d,positioning,quotes}

\tikzset{
  tensor/.style={
    inner sep = 0.055cm,
    shape = circle,
    draw,
    fill
  },
  t/.style={
    inner sep = 0.03cm,
    shape = circle,
    draw,
    fill
  } 
    }

\begin{document}

\title{Emergent (2+1)D topological orders from iterative (1+1)D gauging}

\author{ Jos\'e Garre-Rubio}
\affiliation{\mbox{University of Vienna, Faculty of Mathematics, Oskar-Morgenstern-Platz 1, 1090 Vienna, Austria}}

\begin{abstract}
Gauging introduces gauge fields in order to localize an existing global symmetry, resulting in a dual global symmetry on the gauge fields that can be gauged again. By iterating the gauging process on spin chains with Abelian group symmetries and arranging the gauge fields in a 2D lattice, the local symmetries become the stabilizer of the $XZZX$-code for any Abelian group. 
By twisting the gauging map we obtain new codes that explicitly confine anyons, which violate an odd number of plaquette terms and whose fusion results in mobile dipole excitations.  
Our construction naturally realizes any gapped boundary by taking different quantum phases of the initial (1+1)D globally symmetric system. 
Our method establishes a new route to obtain higher dimensional topological codes from lower ones, to identify their gapped boundaries and their tensor network representations.
\end{abstract}

\maketitle

Gauging is fundamental in the Standard Model to comprehend and unify forces. It transforms a system, promoting its global symmetry to a local symmetry by introducing new degrees of freedom known as gauge fields. While the initial motivation for gauging was Lagrangians supported on the continuum with Lie group symmetries, the gauging of quantum lattice Hamiltonians with finite group symmetries has gained significant attention \cite{Levin_2012,Haegeman14, Williamson16}. 

The power of gauging lies in the fact that it connects very distinct phases of matter, which makes it the standard tool to classify quantum phases and to prove the existence of anomalies \cite{Barkeshli14, thorngren2015higher, Yoshida16}. Since gauging global (1+1)D symmetries results in emergent dual global symmetries (which could be non-invertible for non-Abelian groups \cite{Gaiotto15,Bhardwaj18}) this turns gauging into the source of dualities in (1+1)D \cite{lootens21dualA}. In (2+1)D, the emergent symmetries give rise to a very rich phenomena including 1-form and surface symmetries \cite{delcamp2023higher}. Gauging has also been generalized to other settings beyond on-site global symmetries, including non-on-site global symmetries \cite{Garre23} and higher form symmetries \cite{Yoshida16,Shirley19,moradi2023symmetry,RayhaunWilliamson23}, leading to the creation of fractal phases \cite{Williamson16B,Vijay16}. 

All previous gauging and duality setups relate systems in the same physical dimension. In this work we use gauging to establish a bulk-boundary correspondence: the construction of a $(2+1)$D topologically ordered system (with local symmetries) from  $(1+1)$D globally symmetric systems.

To achieve this, we iteratively gauge the emergent 1D global symmetries of the new gauge fields for finite Abelian groups. Since the corresponding matter fields are not discarded at every step, we arrange them as the horizontal layers of the newly constructed 2D lattice. Unexpectedly, the local symmetries from each gauging, modified by the composition of the subsequent maps, become the stabilizers of the generalization of the $XZZX$-code \cite{XZZX21}(a realization of the toric code \cite{Kitaev03} proposed in \cite{Wen03}) for any Abelian group. By twisting the gauging map by a $2$-cocycle \cite{Thorngren19, Blanik24}, we explicitly confined anyons that now violate an odd number of plaquette terms and whose fusion results in mobile dipoles. 

The different gapped boundaries (and hence the condensable anyons at the boundary) of our construction depend on the quantum phase of the initial (1+1)D globally symmetric system. We show this by establishing a connection between boundary Hamiltonian terms and (1+1)D string order parameters evaluated on the initial system. Such connection illuminates the fact that both settings, gapped boundaries of quantum doubles of $G$ and (1+1)D quantum phases with global symmetries, are classified by the same mathematical object.

Since the gauging operator is a tensor network, our 2D construction inherits that structure, giving rise to a subfamily of projected entangled pair states (PEPS) \cite{Verstraete04} that we refer to as projected entangled pair emergent states (PEPES) that satisfy a different version of the virtual symmetry leading to topological ordered PEPS \cite{Schuch12, Sahinoglu14}.

\textit{\bf Gauging}-- The procedure of gauging maps globally symmetric operators and states $\{ O,\ket{\psi}\}$, to local symmetric ones $\{\hat{O},\ket{\hat{\psi}}\}$ such that it preserves their expectation values: $\bra{\psi}O\ket{\psi} = \bra{\hat{\psi}}\hat{O}\ket{\hat{\psi}}$. For the relevant case of Hamiltonians, gauging preserves the gap and maps ground states to ground states of the gauged Hamiltonian \cite{Williamson16B}.

The map is implemented by a gauging operator $\G_0$ \cite{Haegeman14}, that maps the initial {\it matter} Hilbert space $\H_{0}$ to $\H_{0}\otimes \H_{1}$, introducing new degrees of freedom (dof) supported in $\H_{1}$, called the gauge fields.  
Given a global symmetry of a finite Abelian group $G$ represented as $\bigotimes_i u_g^i$ in $\H_{0}$, where $g\in G$ and $i$ denotes the vertices of a 1D chain, the new introduced Hilbert space is $\H_{1} = \bigotimes_{\hat{i}}\mathbb{C}[G]^{\hat{i}}$, where $\hat{i}$ denotes the edge between $i$ and $i+1$ and $\mathbb{C}[G]= {\rm span}\{ |g\rangle, g\in G \}$. We define in $\mathbb{C}[G]$ the unitary representation of $G$ $\{X_g \}_{g\in G}$ as $X_g\ket{h} = \ket{gh}$ that allows to construct the local symmetry projectors 
$ \P^i = \frac{1}{|G|}\sum\limits_{g\in G} X^{\hat{i}\m}_{\myinv{g}}\otimes u^i_g \otimes X^{\hat{i}}_g $. 
Then the global projector to the local symmetric subspace is $\P=\Pi_i \P^i$ such that the gauging operator is defined by $\G_0 = \mathcal{P} \Big( \bigotimes_{ \hat{i}}|e\rangle_{ \hat{i}} \Big )$, where $e$ denotes the trivial group element and it satisfies
\begin{equation}\label{localsymG0}
\left ( X^{\hat{i}\m,1}_{\myinv{g}} \cdot  u_g^{i,0} \cdot X_g^{\hat{i},1} \right ) \G_0 = \G_0 \quad \forall g \in G, \ \forall i \ , \end{equation}
where $j=0,1$ denotes the action on $\H_j$.

Finally gauged states are given by $\ket{\hat{\psi}} = \G_0 | \psi \rangle$ and gauge operators by $\hat{O}\cdot \G_0 = \G_0 \cdot O$. As an example let us consider $H=\sum_i X_iX_{i\p}-Z_i$ with global symmetry $\bigotimes_i Z_i$ that maps after gauging to $\hat{H}=\sum_i X_iZ_{\hat{i}}X_{i\p}-X_{\hat{i}\m}Z_iX_{\hat{i}}$ with local symmetry $X_{\hat{i}\m}Z_iX_{\hat{i}}$ and an {\it emergent} global symmetry $\bigotimes_{\hat{i}} Z_{\hat{i}}$ only supported on the gauge fields. Importantly, we will show next, this emergent global symmetry is always present after gauging.

\textit{\bf The emergent global dual symmetry}-- 
Let us define the operator $Z_{\hat{g}}= \sum_h {\hat{g}}(h)\ket{h}\bra{h} $, associated to an irrep $\hat{g}:G\to U(1)$ of $G$, satisfying $X_g \cdot Z_{\hat{g}} = {\hat{g}}(\myinv{g})\cdot  Z_{\hat{g}}\cdot X_g $. Then, the global operator $\bigotimes_{\hat{i}} Z_{\hat{g}}^{\hat{i},1}$ commutes with the local symmetry of \eqref{localsymG0}, so it does with $\P^i$, and it is a global symmetry of  $\bigotimes_{ \hat{i}}|e\rangle_{ \hat{i}}$. Therefore, the gauged operators and the gauged states endow the following emergent dual global symmetry:
\begin{equation}\label{emertG1} \bigotimes_{\hat{i}} Z_{\hat{g}}^{\hat{i},1} \cdot  \G_0 = \G_0 \quad \forall \hat{g}\in \hat{G}\ , \end{equation}
where the unitary operators $\{Z_{\hat{g}}\}$ are a representation of the dual group $\hat{G}$ of the irreps. In the Appendix we show that for non-Abelian groups the emergent global symmetry is ${\rm Rep}(G)$ and it comes from the zero gauge flux configuration.

In the literature, gauging also involves decoupling and projecting out the matter, resulting in just a gauge fields with a global symmetry, which can be understood as a duality process like in the example $H\to \hat{H}\to \tilde{H}=\sum_i Z_{\hat{i}}-X_{\hat{i}\m}X_{\hat{i}}$, corresponding to Kramers-Wannier duality.

\textit{\bf Iterative Abelian gauging}-- The emergent global $\hat{G}$ symmetry can be gauged as well. To do so we construct the gauging map $\G_1 : \H_{1}\rightarrow \H_{1}\otimes\H_{2}$ by first introducing $\H_{2}= \bigotimes_i\mathbb{C}[\hat{G}]^{i}$, defining the unitary representation of $\hat{G}$ $\{ X_{\hat{g}} \}_{\hat{g}\in \hat{G}}$ as $X_{\hat{g}}\ket{\hat{h}} = \ket{\hat{g}\hat{h}}$, and then projecting $\bigotimes_{ {i}}|\hat{e}\rangle_{ {i},2}$ onto the local symmetric subspace of $X^{{i},2}_{\myinv{\hat{g}}}\otimes Z^{\hat{i},1}_{\hat{g}} \otimes X^{i,2}_{\hat{g}} $. After composing both gauging maps $ \G_1 \circ \G_0: \H_{0} \rightarrow \H_{0}\otimes \H_{1} \otimes \H_{2} $, the initial local symmetry of $\G_0$, see \eqref{localsymG0}, changes to 
\begin{equation}\label{eq:loc2d1}
\left ( X^{\hat{i}\m,1}_{\myinv{g}} \cdot Z^{{i},2}_{\myinv{g}} \cdot u_g^{i,0} \cdot X_g^{\hat{i},1} \right ) \cdot \G_1 \circ \G_0 = \G_1 \circ \G_0 \ ,
\end{equation}
where $Z_g = \sum_{\hat{h}} {\hat{h}}(g)\ket{\hat{h}}\bra{\hat{h}}$ is a representation of $G$ on $\mathbb{C}[\hat{G}]$ and it satisfies $Z_{{g}} \cdot X_{\hat{g}} = {\hat{g}}({g})\cdot X_{\hat{g}} \cdot Z_{{g}}$. To get Eq. \eqref{eq:loc2d1} we just have to check that  $\G_1 (X^\dagger_g \otimes  X_g) = (X^\dagger_g\otimes Z^\dagger_g \otimes  X_g) \G_1$ which is how two point symmetric correlation functions (of the global symmetry $\hat{G}$) maps through gauging to string order parameters (of the global symmetry ${G}$): \begin{equation}\label{GuageOP} X^{ i,j}_{g} \cdot  X^{\dagger i',j}_g \ 
\xrightarrow[]{\G_1} \  X^{i,j}_{g}\cdot (\prod_{i\le k < i'} Z^{\hat{k},j+1}_{{g}} )\cdot X^{\dagger i',j}_g\ .
\end{equation}
Again, there is an emergent global symmetry of $G$ after gauging with $\G_1$ realized by $Z_g$ acting on $\H_{2}$. 

\begin{figure}[h!]
 \centering
 \includegraphics[scale=1.5]{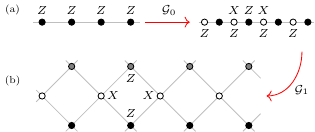}
\caption{Sketch of the iterative gauging for $G=\mathbb{Z}_2$ (a) The global symmetry generated by $Z^{\otimes N}$ is gauged to a local $X\otimes Z\otimes X$ symmetry and a global $Z^{\otimes N}$ on the gauge fields. (b) Emergent local symmetries after applting the second gauging map.}
\label{fig1}
\end{figure}

Therefore, we can iterate the gauging of the emergent global symmetries defining $\G_j : \H_{j}\rightarrow \H_{j}\otimes\H_{j\p}$ and compose $M$ gauging maps:
\begin{equation}\label{defGtotal}
    \G\equiv \G_{M-1} \circ \cdots \circ \G_1 \circ \G_0:\H_{0}\rightarrow \bigotimes_{j=0}^{M} \H_{i} \ ,
 \end{equation}
where $\G_0$ is related to $\G_j\equiv \G_{\rm e}$ by $u_g \leftrightarrow Z_g$ with $j$ even  and $\G_1 = \G_{j}\equiv \G_{\rm o}$ when $j$ odd.

\textit{\bf 2D lattice from iterative 1D gauging}-- 
We place every new Hilbert space $\H_{j+1}$ coming from $\G_j : \H_{j}\rightarrow \H_{j}\otimes\H_{j\p} ,$ on the next layer of a 2D array. Every local Hilbert space $\mathbb{C}[G]$ will be on the vertices $\{i\}$, and $\mathbb{C}[\hat{G}]$ on the edges $\{\hat{i}\}$ (placed between $i$ and $i+1$). This creates a rotated squared lattice with the following symmetries:
\begin{align*} \left ( X^{\hat{i}\m,j}_{\myinv{g}} \cdot Z^{{i},j+1}_{\myinv{g}} \cdot Z_g^{i,j-1} \cdot X_g^{\hat{i},j} \right )\cdot \G = \G \ , j \  {\rm odd} \\
\left ( X^{{i}\m,j}_{\myinv{\hat{g}}} \cdot Z^{\hat{i},j+1}_{\myinv{\hat{g}}} \cdot Z_{\hat{g}}^{\hat{i},j-1} \cdot X_{\hat{g}}^{{i},j} \right )\cdot \G = \G \ ,  j \  {\rm even}
\end{align*}

These local symmetries commute since $[Z_{{g}}^\dagger \otimes X_{{g}} , X^\dagger_{\hat{g}} \otimes Z_{\hat{g}}]= 0$. Remarkably, for $G = \mathbb{Z}_2$ they are the stabilizers of the $XZZX$-code \cite{XZZX21}, which is a different realization of the toric code \cite{Kitaev03,Wen03}.

So our state $\G$ is a common (+1) eigenstate of the aforementioned commuting stabilizer terms which can  be seen as the ground state of the topological code Hamiltonian:
\begin{equation}\label{Hbulk}
H^{\rm Emerg.}_{G,{\rm bulk}} = 
-\!  \sum_{\substack{g \in G\\e\neq g}} \!
\begin{tikzpicture}[baseline=0cm]
\draw[black, rotate around={45:(0,0)},fill=lightgray, opacity=0.5] (-0.4,-0.4) rectangle (0.4,0.4);

\draw[black, rotate around={45:(0,0)},fill=lightgray, opacity=0.5] (-0.55,-0.4)--(-0.4,-0.4)--(-0.4,-0.55);
\draw[black, rotate around={45:(0,0)},fill=lightgray, opacity=0.5] (0.55,0.4)--(0.4,0.4)--(0.4,0.55);
\draw[black, rotate around={45:(0,0)},fill=lightgray, opacity=0.5] (-0.55,0.4)--(-0.4,0.4)--(-0.4,0.55);
\draw[black, rotate around={45:(0,0)},fill=lightgray, opacity=0.5] (0.55,-0.4)--(0.4,-0.4)--(0.4,-0.55);
\node[] at (-0.45,0) {$X^\dagger_g$};
\node[] at (0.5,0) {$X_g$};
\node[] at (0.05,0.45) {$Z^\dagger_g$};
\node[] at (0,-0.45) {$Z_g$};
\end{tikzpicture} 
-\! \sum_{\substack{ \hat{g} \in \hat{G}\\\hat{g}\neq \hat{e}} } \!
\begin{tikzpicture}[baseline=0cm]
\draw[black, rotate around={45:(0,0)},fill=lightgray, opacity=0.5] (-0.55,-0.4)--(-0.4,-0.4)--(-0.4,0.4)--(-0.55,0.4);
\draw[black, rotate around={45:(0,0)},fill=lightgray, opacity=0.5] (-0.4,-0.55)--(-0.4,-0.4)--(0.4,-0.4)--(0.4,-0.55);
\draw[black, rotate around={45:(0,0)},fill=lightgray, opacity=0.5] (0.55,-0.4)--(0.4,-0.4)--(0.4,0.4)--(0.55,0.4);
\draw[black, rotate around={45:(0,0)},fill=lightgray, opacity=0.5] (0.4,0.55)--(0.4,0.4)--(-0.4,0.4)--(-0.4,0.55);

\node[] at (-0.45,0) {$X^\dagger_{\hat{g}}$};
\node[] at (0.45,0) {$X_{\hat{g}}$};
\node[] at (0,0.45) {$Z^\dagger_{\hat{g}}$};
\node[] at (0,-0.45) {$Z_{\hat{g}}$};
\end{tikzpicture} 
\end{equation}
Therefore, we have constructed the generalization of the $XZZX$-code for any Abelian group $G$ by using the emergent symmetries of the concatenation of $(1+1)$D gauging. 

$H^{\rm Emerg.}_{G,{\rm bulk}}$ commutes with $\{ \bigotimes_{\hat{i}} Z^{\hat{i},j}_{\hat{g}}\}_{j\ {\rm odd}}^{\hat{g}}$ and $\{ \bigotimes_{{i}} Z^{{i},j}_{{g}}\}_{j\ {\rm even}}^{g}$, the emergent global symmetries of the ground state $\G$, which correspond to the horizontal logical operators. The vertical logical operators are 
$\{ \bigotimes_{\hat{j}} X^{\hat{i},\hat{j}}_{\hat{g}}\}_{\hat{i}}^{\hat{g}}$ and $\{ \bigotimes_{{j}} X^{{i},j}_{{g}}\}^{g}_i$, that applied to $\G$ generate the $|G|^2$ ground states of $H^{\rm Emerg.}_{G,{\rm bulk}}$.

\begin{figure}[h!]
 \centering
 \includegraphics[scale=1.1]{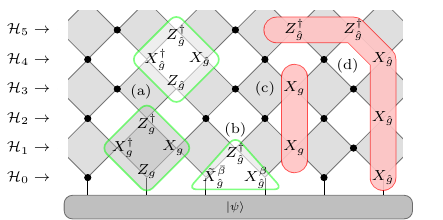}
\caption{(a) Commuting bulk Hamiltonian terms. (b) Generic boundary term. (c) Vertical anyon given by $g\in G$ that could condense on the boundary. (d) Concatenation of horizontal and vertical anyons given by $\hat{g}\in \hat{G}$. }
\label{figops}
\end{figure}

\textit{ \bf Twisting the bulk}--
The gauging map can be twisted, as introduced in  Ref.\cite{Blanik24}, by a 2-cocycle $\alpha\in \mathcal{H}^2[G,U(1)]$. To do so we introduce the $\alpha$ and $\bar{\alpha}$ projective representations ${X}^{{\alpha}}_{{g}}$ and 
$\tilde{X}^{\bar{\alpha}}_{{g}}$ defined by ${X}^{{\alpha}}_{{g}}\ket{h}=\alpha({g},{h}) \ket{{g}{h} }$ and $\tilde{X}^{\bar{\alpha}}_{{g}} \ket{{h} } = \bar{\alpha}({h} \myinv{{g}},{g})\ket{h\myinv{{g}}}$. These two representations commute so we construct $ \P_\alpha = {|G|^{-1}}\sum_{{g}\in {G}} \tilde{X}^{\bar{\alpha}}_{{g}} \otimes Z_{{g}} \otimes X^{{\alpha}}_{{g}} $ and define $\G_\alpha = \prod_i P_{\alpha}^{i} \otimes_{\hat{i}} \ket{e}^{\hat{i}}$. Importantly the twisted gauging operator also realizes the same emergent dual symmetry of $\hat{G}$: $\bigotimes_{\hat{i}} Z_{\hat{g}}^{\hat{i}} \cdot  \G_\alpha = \G_\alpha$ since the operators satisfy $ Z_{\hat{g}} \cdot X^\alpha_{{g}} = {\hat{g}}({g})\cdot  X^\alpha_{{g}} \cdot Z_{\hat{g}}$.

For untwisted gauging maps on odd layers, the emergent Hamiltonian resulting from concatenating $\G_{\rm o}\G_\alpha$, shares the $\hat{G}$-plaquette terms of \eqref{Hbulk}. However, $G$-plaquette terms are now:
\begin{equation}\label{Twisterms}
B_{g}^\alpha =
\begin{tikzpicture}[baseline=-7pt]
\draw[black, rotate around={45:(0,0)},fill=lightgray, opacity=0.5] (-0.4,-0.4) rectangle (0.4,0.4);
\draw[black, rotate around={45:(0,0)},fill=lightgray, opacity=0.5] (-0.55,-0.4)--(-0.4,-0.4)--(-0.4,-0.55);
\draw[black, rotate around={45:(0,0)},fill=lightgray, opacity=0.5] (0.55,0.4)--(0.4,0.4)--(0.4,0.55);
\draw[black, rotate around={45:(0,0)},fill=lightgray, opacity=0.5] (-0.55,0.4)--(-0.4,0.4)--(-0.4,0.55);
\draw[black, rotate around={45:(0,0)},fill=lightgray, opacity=0.5] (0.55,-0.4)--(0.4,-0.4)--(0.4,-0.55);
\node[] at (-0.45,0) {$\tilde{X}^{\bar{\alpha}}_g$};
\node[] at (0.5,0) {$X^\alpha_g$};
\node[] at (0.05,0.45) {$Z^\dagger_g$};
\node[] at (0,-0.45) {$Z_g$};
\end{tikzpicture} 
\Rightarrow
\prod B_{g}^\alpha = \bigotimes_{i,j} Z^{i,j}_{\imath_{g}\alpha} \ ,
\end{equation}
where $\imath_{{g}\alpha}$ (the so-called slant product) belongs to $\hat{G}$ since $\imath_{{g}\alpha}({h}) = \frac{\alpha({g},{h})}{\alpha({h},{g})}$ and where we have used ${X}^{{\alpha}}_{{g}}\cdot \tilde{X}^{\bar{\alpha}}_{{g}} = Z_{\imath_{{g}}\alpha}$. The fact that the product of all $G$-plaquette terms is the product of the horizontal $\hat{G}$-logical operators, Eq. \eqref{Twisterms}, and not the identity has several consequences. 
First, the only logical vertical operators are
$\{ \bigotimes_{\hat{j}} X^{\hat{i},\hat{j}}_{\hat{g}}\}_{\hat{i}}^{\hat{g}}$, since $\{ \bigotimes_{{j}} X^{{i},j}_{{g}}\}^{g}_i$ do not commute with $B_g^\alpha$. Then the former  generate just a $|G|$-fold ground space: the topological order has changed. Second, there are local operators that violate an odd number of plaquettes:
$$
\begin{tikzpicture}
\begin{scope} [rotate=-45]
\draw[lightgray] (-0.3,0.3) rectangle (0.3+0.6,-0.3-0.6);
\draw[black,fill=lightgray, opacity=0.5] (-0.3,-0.3) rectangle (0.3,0.3);
\draw[black,fill=lightgray, opacity=0.5] (-0.3+0.6,-0.3-0.6) rectangle (0.3+0.6,0.3-0.6);
\filldraw[red] (0,0) circle (1.5pt); 
\filldraw[red] (0,-0.6) circle (1.5pt); 
\filldraw[red] (0.6,-0.6) circle (1.5pt); 
\end{scope}
\node[scale=0.9] at (0.1,-0.5) {$X^\alpha_g$};
\begin{scope} [rotate=-45,shift={(-2,-2)}]
\draw[lightgray] (-0.3,0.3) rectangle (0.3+0.6,-0.3-0.6);
\draw[black,fill=lightgray, opacity=0.5] (-0.3,-0.3) rectangle (0.3,0.3);
\draw[black,fill=lightgray, opacity=0.5] (-0.3+0.6,-0.3-0.6) rectangle (0.3+0.6,0.3-0.6);
\filldraw[red] (0,0) circle (1.5pt); 
\filldraw[red] (0.6,0) circle (1.5pt); 
\filldraw[red] (0.6,-0.6) circle (1.5pt); 
\end{scope}
\node[scale=0.9] at (-2.8,-0.5) {$\tilde{X}^{\bar{\alpha}}_g$};
\end{tikzpicture}
$$
These excitations are confined creating a horizontal string whose energy grows with its length. But gluing them together $\tilde{X}^{\bar{\alpha}}_g\otimes {X}^{{\alpha}}_g$, a dipole excitation is created that moves free vertically. One can bend the dipole excitation by acting with $Z_g$ horizontally, leaving an excitation on the corner or splitting the excitation on to right and left. Notice that the dipole commutes with the horizontal $Z_{\hat{g}}$-string excitations: so these two kind of excitations braid trivially. The only remaining anyons are vertical $X_{\hat{g}}$-strings and horizontal $Z_{g}$-strings that braid non-trivially. 
We can either or also twist the odd layers by ${\gamma}\in \mathcal{H}^2[\hat{G},U(1)]$.

\textit{ \bf Boundary conditions}-- We consider periodic boundary conditions (PBC) in the horizontal direction, see Appendix for open boundary conditions (OBC). We also take $u_g = Z_g$ for simplicity. The vertical boundaries correspond to the input Hilbert space $\H_0$ and $\H_{M}$ of $\G$, see Eq.\eqref{defGtotal}. PBC correspond to $\tr_{\H_0=\H_{M}}[\G]$ for $M$ even, creating a square rotated lattice in a torus.

For vertical OBC, the boundaries correspond to globally symmetric states: $ _{M}\bra{\psi'}\G \ket{\psi}_0$. The boundary terms
that commute with the bulk stabilizers can be chosen to be (see Fig. \ref{figops})
\begin{equation}\label{boundterm}
\tilde{X}^{\bar{\beta}}_{{\hat{g}}} \otimes Z^\dagger_{{\hat{g}}} \otimes  X_{{\hat{g}}}^{\beta},  \quad  \beta\in \mathcal{H}^2[\hat{G},U(1)] \  .
\end{equation}
Then if we concatenate $\ell$ of these terms its action on $\ket{\psi}$ through $\G$ is $\tilde{X}^{\bar{\beta}}_{\hat{g}} \otimes Z_{\imath_{\hat{g}}\beta}^{\otimes\ell}  \otimes X^{{\beta}}_{\hat{g}} $. So the term of Eq.\eqref{boundterm} is a symmetry of $\G \ket{\psi}$ only if  $\bra{\psi}\tilde{X}^{\bar{\beta}}_{\hat{g}} \otimes Z_{\imath_{\hat{g}}\beta}^{\otimes\ell}  \otimes X^{{\beta}}_{\hat{g}} \ket{\psi} = 1$. This correspond to  the expectation value of this string order parameter for the global symmetry $\bigotimes_i Z^i_g$, whose value depends on the quantum phase of $\ket{\psi}$ viewed as one of the ground states of a 1D symmetric Hamiltonian $H^{1D}_\psi$. Therefore, the quantum phase of the 1D boundary determines which boundary stabilizer are present and then which anyons condense at the boundary (by  commuting with all Hamiltonian terms in the boundary). 

Remarkably our construction unifies the fact that the mathematical object that classifies both gapped boundaries of quantum double models of $G$ \cite{Kitaev12,Beigi11,Kong17} and globally $G$ symmetric 1D systems \cite{Schuch11,Chen11} is the same (module categories over Vec$_G$ \cite{ostrik01module}).


Let us analyze in detail the case of $\beta=1$ where the boundary terms are ${X}^{\dagger}_{{\hat{g}}} \otimes Z^\dagger_{{\hat{g}}} \otimes  X_{{\hat{g}}}$ so we evaluate $\bra{\psi} X_{\hat{g}}^{\dagger,i}\otimes X^{ i+\ell}_{{\hat{g}}} \ket{\psi}$, a two point symmetric correlation function that characterizes the pattern of symmetry breaking. If ${H} \subseteq {G}$ is the unbroken symmetry group characterizing the quantum phase of $H^{1D}_\psi$, generically $\bra{\psi}(X^i_{\hat{g}}\otimes X^{\dagger i'}_{{\hat{g}}}) \ket{\psi} \neq 0 $ if $\hat{g}(h)=1$ for all $h\in H$. 
We further impose that $\bra{\psi}(X^i_{\hat{g}}\otimes X^{\dagger i'}_{{\hat{g}}}) \ket{\psi} = 1$ which is achieved at the RG fixed point, see the Appendix for an explicit construction. So the boundary symmetries correspond to the elements $\hat{g}$ in the subgroup $res^{\hat{G}}_H \subseteq \hat{G}$, where 
$res^{\hat{G}}_H = \{ \hat{g}\in \hat{G} \ | \ {\hat{g}}(H)=1\}$. Therefore, we can construct the associated boundary Hamiltonian as:
\begin{equation}\label{Hbdry}
H^{H,\beta=1}_{{\rm bdry.}} = - 
\sum_{\substack{\ \hat{e}\neq\hat{g} \in res^{\hat{G}}_H }} 
\begin{tikzpicture}[baseline=+5pt]
\draw[black, rotate around={45:(0,0)},fill=lightgray, opacity=0.5] (0.55,-0.4)--(0.4,-0.4)--(0.4,0.4)--(0.55,0.4);
\draw[black, rotate around={45:(0,0)},fill=lightgray, opacity=0.5] (0.4,0.55)--(0.4,0.4)--(-0.4,0.4)--(-0.4,0.55);
\node[] at (-0.5,0) {$X_{{\hat{g}}}$};
\node[] at (0.5,0) {$X^\dagger_{{\hat{g}}}$};
\node[] at (0,0.5) {$Z_{{\hat{g}}}$};
\end{tikzpicture} 
\ ,
\end{equation}
whose terms are violated only by anyons associated to vertical string operators $\bigotimes_{j\in \ell } X^{\hat{i},j}_g$ ending in the boundary with $g\in G\setminus H$, see Fig. \ref{figops} (c). At the contrary, a vertical anyon of type $g\in H$ condenses at the boundary and anyons indexed by $\hat{g}\in \hat{G}$ (see Fig.\ref{figops} (d)) condense too. If $H=G$, the Hamiltonian of Eq.\eqref{Hbdry} is trivial so no boundary terms can be violated: all anyon can condense. 
If we would have started from a $\hat{G}$ global symmetry, the only anyons violating the boundary would be indexed by $\hat{G}\setminus \hat{H}$. The previous discussion can also be applied to last layer $\ket{\psi'}$ of $ _M\bra{\psi'}\G \ket{\psi}_0$.

\textit{ \bf Tensor network description}-- The gauging operators, $\G_{\rm o,e}$ in odd and even layers, are matrix product operators  constructed from two tensors (see Appendix) of bond dimension $|G|$:
$$ \G_{\rm o,e} =
\begin{tikzpicture}[baseline=-1pt]
\foreach \x in {0,2,4}{
\node[tensor] at (\x,0) {};
\node[scale=0.9] at (\x+0.35,-0.3) {$M_{\rm o,e}$};
\draw[thick] (\x-0.5,0)--(\x+0.5,0);
\draw[thick] (\x,-0.3)--(\x,0)--(\x+0.2,0.2);
}
\foreach \x in {1,3,5}{
\node[tensor] at (\x,0) {};
\node[scale=0.9] at (\x+0.2,-0.3) {$T_{\rm o,e}$};
\draw[thick] (\x-0.5,0)--(\x+0.5,0);
\draw[thick] (\x,0)--(\x,0.3+0);
}
\end{tikzpicture}
\ .$$

Then, the state $\G$ in \eqref{defGtotal} is a projected entangled pair state (PEPS) \cite{Verstraete04} emerging from the concatenation of 1D gauging operators. Subsequently, we dub this subfamily of PEPS as projected entangled pair emergent states (PEPES). The two different tensors, corresponding to the two types of vertices in the rotated squared checkboard lattice of the PEPES (see Fig.\ref{figops}) have the  following symmetries: 
\begin{align*}
\begin{tikzpicture}[baseline=-0.2cm]
\node[scale=0.9] at (-0.25,0.2) {$M_{\rm e}$};
\node[scale=0.9] at (-0.25,-0.3) {$T_{\rm o}$};
\node[tensor] at (0,0) {};
\node[tensor] at (0,-0.5) {};
\draw[thick] (-0.3,0)--(0.3,0);
\draw[thick] (0,-0.2)--(0,0)--(0.2,0.2);
\draw[thick] (0,-0.2)--(0,-0.5);
\draw[thick] (-0.3,-0.5)--(0.3,-0.5);
\end{tikzpicture}
=
\begin{tikzpicture}[baseline=-1pt]
\draw[thick,fill=lightgray,opacity=0.7] (-0.3,-0.3)--(0,0)--(0.3,-0.3);
\draw[thick,fill=lightgray,opacity=0.7] (-0.3,0.3)--(0,0)--(0.3,0.3);
\node[tensor] at (0,0) {};
\draw[thick] (0,0)--(0,0.2);
\end{tikzpicture}
= 
\begin{tikzpicture}[baseline=-1pt]
\node[scale=0.9] at (-0.5,0.3) {$X_{{g}}$};
\node[scale=0.9] at (0.5,0.3) {$X^\dagger_{{g}}$};
\node[scale=0.9] at (-0.55,-0.3) {$Z_{{g}}$};
\node[scale=0.9] at (0.5,-0.3) {$Z^\dagger_{{g}}$};
\draw[thick,fill=lightgray,opacity=0.5] (-0.3,-0.3)--(0,0)--(0.3,-0.3);
\draw[thick,fill=lightgray,opacity=0.5] (-0.3,0.3)--(0,0)--(0.3,0.3);
\node[tensor] at (0,0) {};
\draw[thick] (0,0)--(0,0.2);
\end{tikzpicture}
\begin{tikzpicture}[baseline=-1pt]
\node[scale=1] at (-0.8,0) {$=$};
\node[scale=0.9] at (-0.5,0.3) {$Z^\dagger_{\hat{g}}$};
\node[scale=0.9] at (0.5,0.3) {$Z_{\hat{g}}$};
\draw[thick,fill=lightgray,opacity=0.7] (-0.3,-0.3)--(0,0)--(0.3,-0.3);
\draw[thick,fill=lightgray,opacity=0.7] (-0.3,0.3)--(0,0)--(0.3,0.3);
\node[tensor] at (0,0) {};
\draw[thick] (0,0)--(0,0.2);
\end{tikzpicture}
\begin{tikzpicture}[baseline=-1pt]
\node[scale=1] at (-0.8,0) {$=$};
\node[scale=0.9] at (-0.5,-0.3) {$X^\dagger_{\hat{g}}$};
\node[scale=0.9] at (0.5,-0.3) {$X_{\hat{g}}$};
\draw[thick,fill=lightgray,opacity=0.5] (-0.3,-0.3)--(0,0)--(0.3,-0.3);
\draw[thick,fill=lightgray,opacity=0.5] (-0.3,0.3)--(0,0)--(0.3,0.3);
\node[tensor] at (0,0) {};
\draw[thick] (0,0)--(0,0.2);
\end{tikzpicture}
 , \\
\begin{tikzpicture}[baseline=-0.2cm]
\node[scale=0.9] at (-0.25,0.2) {$M_{\rm o}$};
\node[scale=0.9] at (-0.25,-0.3) {$T_{\rm e}$};
\node[tensor] at (0,0) {};
\node[tensor] at (0,-0.5) {};
\draw[thick] (-0.3,0)--(0.3,0);
\draw[thick] (0,-0.2)--(0,0)--(0.2,0.2);
\draw[thick] (0,-0.2)--(0,-0.5);
\draw[thick] (-0.3,-0.5)--(0.3,-0.5);
\end{tikzpicture}
=
\begin{tikzpicture}[baseline=-1pt]
\draw[thick,fill=lightgray,opacity=0.7] (-0.3,-0.3)--(0,0)--(-0.3,0.3);
\draw[thick,fill=lightgray,opacity=0.7] (0.3,-0.3)--(0,0)--(0.3,0.3);
\node[tensor] at (0,0) {};
\draw[thick] (0,0)--(0,0.2);
\end{tikzpicture}
= 
\begin{tikzpicture}[baseline=-1pt]
\node[scale=0.9] at (-0.5,0.3) {$X_{\hat{g}}$};
\node[scale=0.9] at (0.5,0.3) {$X^\dagger_{\hat{g}}$};
\node[scale=0.9] at (-0.5,-0.3) {$Z_{\hat{g}}$};
\node[scale=0.9] at (0.5,-0.3) {$Z^\dagger_{\hat{g}}$};
\draw[thick, fill=lightgray,opacity=0.7] (-0.3,-0.3)--(0,0)--(-0.3,0.3);
\draw[thick, fill=lightgray,opacity=0.7] (0.3,-0.3)--(0,0)--(0.3,0.3);
\node[tensor] at (0,0) {};
\draw[thick] (0,0)--(0,0.2);
\end{tikzpicture}
\begin{tikzpicture}[baseline=-1pt]
\node[scale=1] at (-0.8,0) {$=$};
\node[scale=0.9] at (-0.5,-0.3) {$X^\dagger_{{g}}$};
\node[scale=0.9] at (0.5,-0.3) {$X_{{g}}$};
\draw[thick,fill=lightgray,opacity=0.7] (-0.3,-0.3)--(0,0)--(-0.3,0.3);
\draw[thick,fill=lightgray,opacity=0.7] (0.3,-0.3)--(0,0)--(0.3,0.3);
\node[tensor] at (0,0) {};
\draw[thick] (0,0)--(0,0.2);
\end{tikzpicture}
\begin{tikzpicture}[baseline=-1pt]
\node[scale=1] at (-0.8,0) {$=$};
\node[scale=0.9] at (-0.5,0.3) {$Z^\dagger_{{g}}$};
\node[scale=0.9] at (0.5,0.3) {$Z_{{g}}$};
\draw[thick,fill=lightgray,opacity=0.7] (-0.3,-0.3)--(0,0)--(-0.3,0.3);
\draw[thick,fill=lightgray,opacity=0.7] (0.3,-0.3)--(0,0)--(0.3,0.3);
\node[tensor] at (0,0) {};
\draw[thick] (0,0)--(0,0.2);
\end{tikzpicture}
.
\end{align*}
The first three relations of each tensor correspond to the virtual $G$ and $\hat{G}$-invariance characterizing 2D topological order in PEPS \cite{Schuch12, Sahinoglu14} and the last relations makes compatible those symmetries. Note that the virtual loop symmetries propagates only in the horizontal direction. 

\textit{ \bf Another view on our construction}-- 
We can interpret $\G$ as the projection of stacked layers of 1D product states. By using that $\G_j= \P_j \bigotimes_i \ket{e}^{i}_j$ we can write
$$ 
\G = (\prod_{j=1} \P_{2j} \cdot \P_{2j-1})\P_0 \left ( \ket{\psi}\bigotimes_{j}
\otimes_{\hat{i}}\ket{e}^{\hat{i}}_{2j}\otimes_{i}\ket{\hat{e}}^{{i}}_{2j-1}\right ) ,
 $$
where $\otimes_{\hat{i}}\ket{e}^{\hat{i}}$ is locally invariant under any $Z^{\hat{i}}_g$ and the projectors $P_{2j}$ and $\P_{2j-1}$ do not commute. The two previous properties differ from the common approach of creating topologically ordered models from stacking lower dimensional ones (where the coupling generally commutes) \cite{Kane02,Brown11, Neupert14, WilliamsonDevakul21,tantivasadakarn22,DomCheng23}.

\textit{\bf Conclusions}--In this work we have established a bulk-boundary correspondence between 1D global symmetric systems and 2D topologically ordered models. We do so by sequentially gauging the emergent 1D global symmetries that maps the local 1D symmetries after gauging to 2D plaquette operators. As a result we obtain a family  of 2D Hamiltonians:
$$ H^G_{\rm Emerg.} = H_{\rm bulk}^{\alpha, \gamma} + H^{[\ket{\psi}],\beta}_{{\rm bdry.}}\ . $$
This family covers the generalization of the $XZZX$-code for any Abelian group $G$. Also, these Hamiltonians are able to realize novel anyon confinement phenomena where there are local excitations violating $3$ plaquette terms. Moreover, the boundary terms are given by the quantum phase of the 1D Hamiltonian of $\ket{\psi}$ and determines which anyons condense at the boundary. Such connection illuminates the fact that both settings are classified by the  same mathematical object.

The questions of how our construction can be generalized to non-Abelian topological orders and its connection with Ref.\cite{Liu23} remain open. Finally, in a forthcoming paper \cite{BramPepe}, we will present the emergence of non-trivial (3+1)D phases from the gauging of (2+1)D symmetries.

\section*{Acknowledgements}

I thank David Blanik for the discussion that inspired this work, Bram Vancraeynest-De Cuiper, Aleksander Kubicki, Anasuya Lyons and Isabel Miranda for their helpful comments on the manuscript and Dominic Williamson for pointing out Ref.\cite{Liu23}. 
\bibliography{bibliography.bib}

\appendix
\section{Dual ${\rm Rep}(G)$ symmetry}

When gauging non-Abelian groups we need to define
the left and right regular representations acting on $\mathbb{C}[G]$: $L_g|h\rangle = |gh\rangle$ and $R_g|h\rangle = |h\myinv{g}\rangle$. So that we define $ \P^i = \frac{1}{|G|}\sum\limits_{g\in G} R^{\hat{i}\m}_{{g}}\otimes u^i_g \otimes L^{\hat{i}}_g $, satisfying that $ [\P^i,\P^{i+1}]=0$ since $[L_h,R_g]=0$. 
Let us write down again the gauging map:
$$\G_0 = \sum_{\{g_i\}}  \cdots u_{g_{i\m}}\otimes \ket{g_{i\m} \myinv{g_{i}}}\otimes  u_{g_i}\otimes \ket{g_i \myinv{g_{i\p}}} \otimes u_{g_{i\p}}  \cdots \ ,$$
where we can see that the 'product' of the gauge field configurations are always the identity element: $\prod_{i=1}^n g_i \myinv{g_{i\p}} = e$. We interpret this as the gauging map creating zero total flux configurations.  
Then, measuring this zero flux with a  {\it charge}, associated to an irreducible representation (irrep) of $G$, will give a trivial action. This means that there is a global symmetry from the dual of $G$, i.e. $Rep(G)$, on the gauge fields. $Rep(G)$ is not a group when $G$ is not Abelian, it is a {\it fusion category} whose simple objects are the irreps of $G$, denoted here as $\sigma, \rho,\gamma,$ etc. The multiplication, a.k.a fusion, is given by $\sigma \cdot \rho = \sum_\gamma N_{\sigma,\rho}^\gamma \gamma $ which accounts for the tensor product of irreps and its direct sum decomposition: $\sigma \otimes \rho \simeq \bigoplus_\gamma \id_{N_{\sigma,\rho}^\gamma} \otimes\gamma $. 

Given an irrep $\sigma$ of $G$, a representation of the fusion category $Rep(G)$ on the Hilbert space $\H_{1}=\mathbb{C}[G]^{\otimes n}$ can be constructed using the following matrix product operators (appeared also in Ref.\cite{fechisin23}):
\begin{equation}\label{Opirrep}
    \Gamma_\sigma =\sum_{\{g_i\}} \chi_\sigma\left ( \prod_{i=1}^n g_i\right )\ket{g_1,\cdots,g_n}\bra{g_1,\cdots,g_n}\ ,
\end{equation}
where $\chi_\sigma:G\to U(1)$ is the irrep character of $\sigma$. Finally the operators satisfy:
$$\Gamma_\sigma \cdot \Gamma_\rho = \sum_\gamma N_{\sigma,\rho}^\gamma \Gamma_\gamma \ ,$$
where the relation can be seen by using that $\chi_\sigma\cdot \chi_\rho= \chi_{\sigma \otimes \rho} = 
\sum_\gamma N_{\sigma,\rho}^\gamma \chi_{\gamma} $. Therefore, the dual global symmetry on the gauge fields reads:
\begin{equation}
    (\id_{\H_{0}}\otimes \Gamma_\sigma|_{\H_{1}} )\cdot \G_0 = {d_\sigma}\G_0 \quad \forall \sigma \in Rep(G) .
\end{equation}
Notice that if the total flux is not zero, say $g_{\rm total}\in G$, then the action of $\Gamma_\sigma$ results in $\chi_\sigma(g_{\rm total})$. Finally this dual global symmetry does not commute with the local symmetry:
$$(R_g\otimes u^{[i]}_g \otimes L_g) \cdot \Gamma_\sigma = \Gamma_\sigma \cdot (R^\dagger_g\otimes u^{[i]}_g \otimes L^\dagger_g) \ .$$

\section{Boundary states of symmetry broken phases}

We consider the different symmetry-broken phases of $G$, indexed by the unbroken symmetry group $H \subseteq G$. For the sake of presentation we choose the global symmetry to be $X_g^{\otimes n}$. Then a representation of the ground space of a Hamiltonian in a $H$ symmetry unbroken phase can be given by ${\rm span }\{ ( X_g \ket{+}_H )^{\otimes n}, \ g\in G \}$, where $\ket{+}_H = \frac{1}{\sqrt{|H|}}\sum_{h \in H} \ket{h}$. Then, within this ground space we will choose as representative the following symmetric ground state:
$$
\ket{\psi}_H=\frac{1}{\sqrt{|G|}}\sum_{g\in G} (X_g\ket{+}_H)^{\otimes n} \ .
$$ 
For example, the maximally symmetry-broken phase and the disordered phase are given by $\ket{\psi}_{\mathbb{Z}_1} =\frac{1}{\sqrt{|G|}} \sum_{g\in G} \ket{g}^{\otimes n}$ and $\ket{\psi}_G = \ket{+}_G^{\otimes n}$ respectively. Then it is easy to check that 
$$(Z_{\hat{g}})^{[i]}\otimes( Z^\dagger_{\hat{g}})^{[j]} \ket{\psi}_H = \ket{\psi}_H \Leftrightarrow {\hat{g}}(H)=1 \ . 
$$

\section{Tensor network construction}

The gauging map for $(1+1)$D systems has a very  simple form as a tensor network. It is a matrix product operator with bond dimension $|G|$:
$$ \G_0 =
\begin{tikzpicture}[baseline=-1pt]
\foreach \x in {0,2,4}{
\node[tensor] at (\x,0) {};
\draw[thick] (\x-0.5,0)--(\x+0.5,0);
\draw[thick] (\x,-0.3)--(\x,0)--(\x+0.2,0.2);
}
\foreach \x in {1,3,5}{
\foreach \y in {0}{
\node[tensor,blue] at (\x,\y) {};
\draw[thick,blue] (\x-0.5,\y)--(\x+0.5,\y);
\draw[thick,blue] (\x,\y)--(\x,0.3+\y);
}}
\end{tikzpicture}
\ ,
$$
where the tensors involved and their symmetries are:
\begin{align*}
\tilde{M}=
\begin{tikzpicture}[baseline=-1pt]
\node[tensor] at (0,0) {};
\draw[thick] (-0.3,0)--(0.3,0);
\draw[thick] (0,-0.3)--(0,0)--(0.2,0.2);
\end{tikzpicture}
& = \sum_g u_g\otimes \ket{g}\bra{g} \ ,
\\
T_{\rm e}=
\begin{tikzpicture}
\node[tensor,blue] at (0,0) {};
\draw[thick,blue] (-0.3,0)--(0.3,0);
\draw[thick,blue] (0,0.3)--(0,0);
\end{tikzpicture}
& = \frac{1}{|G|}\sum_{g,h} \ket{g\myinv{h}}\otimes \ket{g}\bra{h}
\\
&=
\begin{tikzpicture}[baseline=0.2cm]
\node[tensor,blue] at (0,0) {};
\draw[thick,blue] (-0.3,0)--(0.3,0);
\draw[thick,blue] (0,0.3)--(0,0);
\node[scale=0.9] at (0.55,0) {$Z^\dagger_{\hat{g}}$};
\node[scale=0.9] at (-0.55,0) {$Z_{\hat{g}}$};
\node[scale=0.9] at (0,0.55) {$Z^\dagger_{\hat{g}}$};
\end{tikzpicture}
=
\begin{tikzpicture}[baseline=0.1cm]
\node[tensor,blue] at (0,0) {};
\draw[thick,blue] (-0.3,0)--(0.3,0);
\draw[thick,blue] (0,0.3)--(0,0);
\node[scale=0.9] at (0.55,0) {$X^\dagger_{{g}}$};
\node[scale=0.9] at (-0.55,0) {$X_{{g}}$};
\end{tikzpicture}
\  .
\end{align*}
Notice that $\G_0$ is related to $\G_2$ by $u_g \leftrightarrow Z_g$ and $\G_2 = \G_{2i}$ for all $i\in \mathbb{N}$. Similarly $\G_1 = \G_{2i+1}$ for all $i \in  \mathbb{N} $ then, by defining the following tensors (with their corresponding symmetries):
\begin{align*}
M_{\rm o}=
\begin{tikzpicture}[baseline=-1pt]
\node[tensor,red] at (0,0) {};
\draw[thick,red] (-0.3,0)--(0.3,0);
\draw[thick,red] (0,-0.3)--(0,0)--(0.2,0.2);
\end{tikzpicture}
& = \sum_{\hat{g}} Z_{\hat{g}}\otimes \ket{\hat{g}}\bra{\hat{g}}
=
\begin{tikzpicture}[baseline=-0.1cm]
\node[tensor,red] at (0,0) {};
\draw[thick,red] (-0.3,0)--(0.3,0);
\draw[thick,red] (0,-0.3)--(0,0)--(0.2,0.2);
\node[scale=0.9] at (0.2,0.5) {$Z_{\hat{g}}$};
\node[scale=0.9] at (0,-0.55) {$Z^\dagger_{\hat{g}}$};
\end{tikzpicture} 
=
\begin{tikzpicture}[baseline=-0.1cm]
\node[tensor,red] at (0,0) {};
\draw[thick,red] (-0.3,0)--(0.3,0);
\draw[thick,red] (0,-0.3)--(0,0)--(0.2,0.2);
\node[scale=0.9] at (0,-0.55) {$Z_{\hat{h}}$};
\node[scale=0.9] at (0.55,0) {$X^\dagger_{\hat{h}}$};
\node[scale=0.9] at (-0.55,0) {$X_{\hat{h}}$};
\end{tikzpicture} 
\ ,
\\
T_{\rm o}=
\begin{tikzpicture}
\node[tensor,red] at (0,0) {};
\draw[thick,red] (-0.3,0)--(0.3,0);
\draw[thick,red] (0,0.3)--(0,0);
\end{tikzpicture}
& = \frac{1}{|G|}\sum_{\hat{g},\hat{h}} \ket{\hat{g}\myinv{\hat{h}}}\otimes \ket{\hat{g}}\bra{\hat{h}}
\\
&=
\begin{tikzpicture}[baseline=0.2cm]
\node[tensor,red] at (0,0) {};
\draw[thick,red] (-0.3,0)--(0.3,0);
\draw[thick,red] (0,0.3)--(0,0);
\node[scale=0.9] at (0.55,0) {$Z^\dagger_{{g}}$};
\node[scale=0.9] at (-0.55,0) {$Z_{{g}}$};
\node[scale=0.9] at (0,0.55) {$Z^\dagger_{{g}}$};
\end{tikzpicture}
=
\begin{tikzpicture}[baseline=0.1cm]
\node[tensor,red] at (0,0) {};
\draw[thick,red] (-0.3,0)--(0.3,0);
\draw[thick,red] (0,0.3)--(0,0);
\node[scale=0.9] at (0.55,0) {$X^\dagger_{\hat{g}}$};
\node[scale=0.9] at (-0.55,0) {$X_{\hat{g}}$};
\end{tikzpicture}
\ ,
\\
M_{\rm e}=
\begin{tikzpicture}[baseline=-1pt]
\node[tensor,blue] at (0,0) {};
\draw[thick,blue] (-0.3,0)--(0.3,0);
\draw[thick,blue] (0,-0.3)--(0,0)--(0.2,0.2);
\end{tikzpicture}
&= \sum_g Z_g\otimes \ket{g}\bra{g}
=
\begin{tikzpicture}[baseline=-0.1cm]
\node[tensor,blue] at (0,0) {};
\draw[thick,blue] (-0.3,0)--(0.3,0);
\draw[thick,blue] (0,-0.3)--(0,0)--(0.2,0.2);
\node[scale=0.9] at (0.2,0.5) {$Z_{{g}}$};
\node[scale=0.9] at (0,-0.55) {$Z^\dagger_{{g}}$};
\end{tikzpicture} 
=
\begin{tikzpicture}[baseline=-0.1cm]
\node[tensor,blue] at (0,0) {};
\draw[thick,blue] (-0.3,0)--(0.3,0);
\draw[thick,blue] (0,-0.3)--(0,0)--(0.2,0.2);
\node[scale=0.9] at (0,-0.55) {$Z_{{h}}$};
\node[scale=0.9] at (0.55,0) {$X^\dagger_{{h}}$};
\node[scale=0.9] at (-0.55,0) {$X_{{h}}$};
\end{tikzpicture}
\ .
\end{align*}
By using the previous defined tensors we can write our 2D construction of iterative gauging as follows:
\begin{equation}\label{TNG}
\G = \!
\begin{tikzpicture}[baseline=+1cm]
\foreach \x in {0,2,4}{
\node[tensor] at (\x,0) {};
\draw[thick] (\x-0.5,0)--(\x+0.5,0);
\draw[thick] (\x,-0.3)--(\x,0)--(\x+0.2,0.2);
\foreach \y in {1.2}{
\node[tensor,blue] at (\x,\y) {};
\draw[thick,blue] (\x-0.5,\y)--(\x+0.5,\y);
\draw[thick,blue] (\x,-0.3+\y)--(\x,\y)--(\x+0.2,0.2+\y);
}}
\foreach \x in {1,3,5}{
\foreach \y in {0,1.2}{
\node[tensor,blue] at (\x,\y) {};
\draw[thick,blue] (\x-0.5,\y)--(\x+0.5,\y);
\draw[thick,blue] (\x,\y)--(\x,0.3+\y);
}}
\foreach \x in {1,3,5}{
\foreach \y in {0.6,1.8}{
\node[tensor,red] at (\x,\y) {};
\draw[thick,red] (\x-0.5,\y)--(\x+0.5,\y);
\draw[thick,red] (\x,-0.3+\y)--(\x,\y)--(\x+0.2,0.2+\y);
}}
\foreach \x in {0,2,4}{
\foreach \y in {0.6,1.8}{
\node[tensor,red] at (\x,\y) {};
\draw[thick,red] (\x-0.5,\y)--(\x+0.5,\y);
\draw[thick,red] (\x,\y)--(\x,0.3+\y);
}}
\end{tikzpicture}
\ .
\end{equation}

The resulting state is a projected entangled pair state (PEPS) emerging from the concatenation of 1D gauging operators. For that reason, we name the family of tensor network states constructed in this way, a subfamily of PEPS, projected entangled pair emergent states (PEPES). 

The rotated square lattice shape of our 2D construction is better appreciated if we define the following tensors, by appropriately blocking the ones corresponding to the gauging maps:
\begin{align*}
\begin{tikzpicture}[baseline=-0.2cm]
\node[tensor,red] at (0,0) {};
\draw[thick,red] (-0.3,0)--(0.3,0);
\draw[thick,red] (0,-0.2)--(0,0)--(0.2,0.2);
\draw[thick,blue] (0,-0.2)--(0,-0.4);
\node[tensor,blue] at (0,-0.4) {};
\draw[thick,blue] (-0.3,-0.4)--(0.3,-0.4);
\end{tikzpicture}
& =
\begin{tikzpicture}[baseline=-1pt]
\draw[thick,blue] (-0.3,-0.3)--(0,0)--(0.3,-0.3);
\node[tensor,red] at (0,0) {};
\draw[thick,red] (-0.3,0.3)--(0,0)--(0.3,0.3);
\draw[thick,red] (0,0)--(0,0.2);
\end{tikzpicture}
=
\begin{tikzpicture}[baseline=-1pt]
\node[scale=0.9] at (-0.55,0.3) {$X_{\hat{g}}$};
\node[scale=0.9] at (0.55,0.3) {$X^\dagger_{\hat{g}}$};
\node[scale=0.9] at (-0.55,-0.3) {$Z^\dagger_{\hat{g}}$};
\node[scale=0.9] at (0.55,-0.3) {$Z_{\hat{g}}$};
\draw[thick,blue] (-0.3,-0.3)--(0,0)--(0.3,-0.3);
\node[tensor,red] at (0,0) {};
\draw[thick,red] (-0.3,0.3)--(0,0)--(0.3,0.3);
\draw[thick,red] (0,0)--(0,0.2);
\end{tikzpicture}
\ ,
\\
\begin{tikzpicture}[baseline=-0.2cm]
\node[tensor,blue] at (0,0) {};
\draw[thick,blue] (-0.3,0)--(0.3,0);
\draw[thick,blue] (0,-0.2)--(0,0)--(0.2,0.2);
\draw[thick,red] (0,-0.2)--(0,-0.4);
\node[tensor,red] at (0,-0.4) {};
\draw[thick,red] (-0.3,-0.4)--(0.3,-0.4);
\end{tikzpicture}
& =
\begin{tikzpicture}[baseline=-1pt]
\draw[thick,red] (-0.3,-0.3)--(0,0)--(0.3,-0.3);
\node[tensor,blue] at (0,0) {};
\draw[thick,blue] (-0.3,0.3)--(0,0)--(0.3,0.3);
\draw[thick,blue] (0,0)--(0,0.2);
\end{tikzpicture}
=
\begin{tikzpicture}[baseline=-1pt]
\node[scale=0.9] at (-0.55,0.3) {$X_{{g}}$};
\node[scale=0.9] at (0.55,0.3) {$X^\dagger_{{g}}$};
\node[scale=0.9] at (-0.55,-0.3) {$Z^\dagger_{{g}}$};
\node[scale=0.9] at (0.55,-0.3) {$Z_{{g}}$};
\draw[thick,red] (-0.3,-0.3)--(0,0)--(0.3,-0.3);
\node[tensor,blue] at (0,0) {};
\draw[thick,blue] (-0.3,0.3)--(0,0)--(0.3,0.3);
\draw[thick,blue] (0,0)--(0,0.2);
\end{tikzpicture}
\ ,
\end{align*}
Importantly the virtual $G$-symmetry displayed by these blocked tensors is the symmetry that has been identified as the source of topological order in 2D PEPS \cite{Schuch12, Sahinoglu14}.

We now move our study to the horizontal boundaries which  correspond to the boundary conditions on the 1D gauging maps. The simplest case is to consider periodic boundary conditions where the gauging map reads:
\begin{align}\label{GauPBC}
\G^p_0 & = \sum_{\{g_i\}}   \ket{g_n\myinv{g}_1} \otimes u_{g_1}\otimes \ket{g_{1} \myinv{g_{2}}} \cdots \ket{g_{n\m}\myinv{g}_n} \otimes u_{g_n} \\
& =
\begin{tikzpicture}[baseline=-1pt]
\foreach \x in {0,4}{
\node[tensor] at (\x,0) {};
\draw[thick] (\x-0.5,0)--(\x+0.5,0);
\draw[thick] (\x,-0.3)--(\x,0)--(\x+0.2,0.2);
}
\node[] at (2,0) {$\cdots$};
\foreach \x in {-1,1,3}{
\foreach \y in {0}{
\node[tensor,blue] at (\x,\y) {};
\draw[thick,blue] (\x-0.5,\y)--(\x+0.5,\y);
\draw[thick,blue] (\x,\y)--(\x,0.3+\y);
}}
\draw[thick,blue,rounded corners] (-1,0)--++(-0.7,0)--++(0,-0.45)--++ (3.5,0);
\draw[thick,rounded corners] (4,0)--++(0.7,0)--++(0,-0.45)--++ (-3.5,0);
\end{tikzpicture} \notag
\ .
\end{align}
If we compose the gauging maps with periodic  boundary conditions, the resulting PEPES is arranged in a rotated square lattice on a cylinder. For open  boundary conditions we get
\begin{align}\label{GauOBC}
\G^o_0 & =  \sum_{\{g_i\}}   \ket{\myinv{g}_1} \otimes u_{g_1}\otimes \ket{g_{1} \myinv{g_{2}}} \cdots\ket{g_{n \m} \myinv{g_{n}}}\otimes  u_{g_n}\otimes   \ket{g_n} \\
& = 
\begin{tikzpicture}[baseline=-1pt]
\foreach \x in {0,3}{
\node[tensor] at (\x,0) {};
\draw[thick] (\x-0.5,0)--(\x+0.5,0);
\draw[thick] (\x,-0.3)--(\x,0)--(\x+0.2,0.2);
}
\node[] at (2,0) {$\cdots$};
\foreach \x in {-1,1,4}{
\foreach \y in {0}{
\node[tensor,blue] at (\x,\y) {};
\draw[thick,blue] (\x-0.5,\y)--(\x+0.5,\y);
\draw[thick,blue] (\x,\y)--(\x,0.3+\y);
}}
\node[tensor,blue,label=below:$\bra{e}$] at (-1.5,0) {};
\node[tensor,blue,label=below:$\ket{e}$] at (4.5,0) {};
\end{tikzpicture} \notag
\ .
\end{align}

An important point in the open boundary condition case is the fact that the number of d.o.f. introduced (the gauge  fields) exceed the initial number of sites by one. Then, if we compose gauging maps with open  boundary conditions, we get an isosceles trapezoid-shape PEPES. For example, starting with 2 sites and composing 3 times, $\G^o_2\circ \G^o_1 \circ \G^o_0$, we obtain:

$$
\begin{tikzpicture}[baseline=-1pt,scale=0.85]
\foreach \x in {0,2}{
\node[tensor] at (\x,0) {};
\draw[thick] (\x-0.5,0)--(\x+0.5,0);
\draw[thick] (\x,-0.3)--(\x,0)--(\x+0.2,0.2);}
\foreach \x in {-1,1,3}{
\foreach \y in {0}{
\node[tensor,blue] at (\x,\y) {};
\draw[thick,blue] (\x-0.5,\y)--(\x+0.5,\y);
\draw[thick,blue] (\x,\y)--(\x,0.3+\y);
}}
\node[tensor,blue,label=below:$\bra{e}$] at (-1.5,0) {};
\node[tensor,blue,label=below:$\ket{e}$] at (3.5,0) {};
 \foreach \x in {-1,1,3}{
\foreach \y in {0.6}{
\node[tensor,red] at (\x,\y) {};
\draw[thick,red] (\x-0.5,\y)--(\x+0.5,\y);
\draw[thick,red] (\x,-0.3+\y)--(\x,\y)--(\x+0.2,0.2+\y);
}}
\foreach \x in {-2,0,2,4}{
\foreach \y in {0.6}{
\node[tensor,red] at (\x,\y) {};
\draw[thick,red] (\x-0.5,\y)--(\x+0.5,\y);
\draw[thick,red] (\x,\y)--(\x,0.3+\y);
}}
\node[tensor,red,label=below:$\bra{\hat{e}}$] at (-2.5,0.6) {};
\node[tensor,red,label=below:$\ket{\hat{e}}$] at (4.5,0.6) {};
\foreach \x in {-2,0,2,4}{
\foreach \y in {1.2}{
\node[tensor,blue] at (\x,\y) {};
\draw[thick,blue] (\x-0.5,\y)--(\x+0.5,\y);
\draw[thick,blue] (\x,-0.3+\y)--(\x,\y)--(\x+0.2,\y+0.2);}}
\foreach \x in {-3,-1,1,3,5}{
\foreach \y in {1.2}{
\node[tensor,blue] at (\x,\y) {};
\draw[thick,blue] (\x-0.5,\y)--(\x+0.5,\y);
\draw[thick,blue] (\x,\y)--(\x,0.3+\y);
}}
\node[tensor,blue,label=below:$\bra{e}$] at (-3.5,1.2) {};
\node[tensor,blue,label=below:$\ket{e}$] at (5.5,1.2) {};
\end{tikzpicture}
$$

We note that the previous local  boundary symmetries are broken, like $Z_g|_2 \otimes X_g|_1$ or $Z_{\hat{g}}|_3 \otimes X_{\hat{g}}|_2$. This means that any anyon ending on the boundary will not violate the (non-existing) boundary terms so they can condense at the boundary. One way to avoid this shape when using open boundary conditions is to compose the previous construction with the adjoint, that is, 
$$(\G^o_0)^\dagger\circ(\G^o_{1})^\dagger\circ \cdots \circ (\G^o_{N-1})^\dagger \circ\G^o_N\circ \cdots \circ \G^o_1 \circ \G^o_0 , $$
which results in a square lattice (rotated square lattice on a rotated square). 

\section{ From 0D to 1D}
Consider a 0D state, a single site state, with a global (also local) symmetry of $G$: $u_g\ket{\psi} = \ket{\psi}$ i.e. a state  that transforms trivially under $u_g$. We can gauge this by doing the analogous  procedure of 1D resulting in:
\begin{align*} 
\ket{\hat{\psi}} &  = \frac{1}{|G|}\sum_g R_g\otimes u_g \otimes L_g (\ket{0}\ket{\psi} \ket{0}) \\
& =  \frac{1}{|G|}\sum_g \ket{\myinv{g}}\ket{\psi} \ket{g} \ . 
\end{align*}
This shows that $\ket{\hat{\psi}}$ has a global $G\times \hat{G}$ symmetry generated by $(\id \otimes u_g \otimes \id) \times (Z_{\hat{g}}\otimes \id \otimes Z_{\hat{g}}) $ since matter and gauge fields are decoupled (a consequence of gauging a local symmetry). We can now gauge the subgroup $G$ again, which corresponds to a partial gauging, getting a new decoupled pair of sites. If we repeat this process $N$-times, we obtain the following state:
$$ \ket{\psi} \bigotimes_{i=1,\cdots, N} \ket{\omega}_{G}^{[-i,i]},\quad \ket{\omega}_{G}= \frac{1}{|G|}\sum_g \ket{\myinv{g},g} \ .$$
This state is a decoupled long-range entangled state with similar entanglement properties as the rainbow state \cite{Vitagliano10,Ramirez14}.

\end{document}